\documentclass[showpacs,amsmath,amssymb,aps,prl,twocolumn]{revtex4-1}

\usepackage{graphicx}
\usepackage{dcolumn}
\usepackage{bm}

\begin{document}

\title{Temperature-dependent classical phonons from efficient
  non-dynamical simulations}

\author{Mathias P. Ljungberg and Jorge \'I\~niguez}

\affiliation{Institut de Ci\`encia de Materials de Barcelona
  (ICMAB-CSIC), Campus UAB, 08193 Bellaterra, Spain}

\begin{abstract}
We present a method to calculate classical lattice-dynamical
quantities, like the temperature-dependent vibrational spectrum, from
simulations that do not require an explicit solution of the time
evolution. We start from the moment expansion of the relevant
time-correlation function for a many-body system, and show that it can
be conveniently rewritten by using a basis in which the low-order
moments are diagonal. This allows us to approximate the main spectral
features (i.e., position and width of the phonon peaks) from thermal
averages available from any statistical simulation. We illustrate our
method with an application to a model system that presents a
structural transition and strongly temperature-dependent phonons. Our
theory also clarifies the status of previous heuristic schemes to
estimate phonon frequencies.
\end{abstract}


 

\maketitle

Is it possible to compute any equilibrium property of a material from
appropriate thermal averages? From an atomistic simulation
perspective, an affirmative answer to this question implies that, if
we can characterize the configuration space accessible at a
temperature $T$ -- as can be done e.g. with Monte Carlo (MC) methods
--, we can also calculate any quantity of interest. In particular,
time-correlation functions would become available, which would allow
us to compute non-trivial properties like $T$-dependent phonon spectra
or lattice thermal-transport coefficients without explicitly solving
the equations of motion. Such a {\em non-dynamical} approach would
have many advantages: as in MC schemes, fixing $T$ would become
trivial, and we would avoid the problems associated with the use of
thermostats in molecular-dynamics (MD) simulations; there would be no
need for very long MD runs to access low-frequency phenomena, etc.

In the particular case of the vibrational spectrum, several heuristic
schemes have been proposed to realize such a goal. For example,
Hellman {\sl et al}. \cite{hellman11} compute phonons at finite $T$
from an {\em effective} dynamical matrix that they postulate and
obtain in a computationally efficient way; a similar effective
potential is used in Refs.~\onlinecite{brooks95} and
\onlinecite{wheeler03} to capture anharmonic effects in the
vibrational spectrum; others have used the harmonic relation between a
normal-mode frequency and the amplitude of its oscillations to compute
dynamical properties even from powder diffraction
data~\cite{book-dove05,goodwin04,goodwin05}. While such methods are
valuable, we have to resort to earlier works to find a rigorous and
general treatment. As recognized in a variety of fields
\cite{vanvleck48,degennes59,mori65}, a time-correlation function can
be obtained from the knowledge of its moments, which are {\em static}
quantities that can be computed from efficient statistical
simulations. In the context of lattice-dynamical studies, the
moment-based approach introduced by Mori \cite{mori65,balucani03} has
been used to investigate a number of simple systems at both the
quantum and classical levels
\cite{cuccoli92,cuccoli93,cowley94,macchi96,balucani03}; however, it
has failed to gain popularity in the field of classical MD. As far as
we can see, this is probably because (1) we lack a general scheme to
tackle arbitrarily complex materials with many atoms in the cell and
(2) there has not been enough work to identify the simplest
approximations that may render useful results (i.e., reliable phonon
frequencies and possibly peak widths). Here we present our own
derivation of a moment-based formalism for the classical case,
remedying the mentioned deficiencies. The resulting theory allows us
to clarify the status of the heuristic methods in the literature and,
in our opinion, should become a standard tool in the field of
classical simulations.

{\sl General formalism}.-- We define the classical correlation
function of quantities $A_i$ and $B_j$ as
\begin{equation}
C^{AB}_{ij}(t) = \langle A_{i}(0) B_{j}(t) \rangle =
\int_{-\infty}^{\infty} C^{AB}_{ij}(\omega)\, e^{i\omega t} \, d\omega
\end{equation}
where $\langle ... \rangle$ indicates thermal averaging and
$C^{AB}_{ij}(\omega)$ is the corresponding {\em spectral
  function}. Below we will identify $A_{i}$ and $B_{j}$ with atomic
positions or velocities, $i$ and $j$ being composite indices that
label an atom and a direction in space. Note that $C^{AB}_{ij}(t)$ is
real, which implies $C^{AB}_{ij}(-\omega) =
[C^{AB}_{ij}(\omega)]^{*}$. For simplicity, we will work with the real
part $\widetilde{C}^{AA}_{ij}(\omega)$ = $\Re[C^{AA}_{ij}(\omega)]$,
as this will contain the information about the vibrational
spectrum~\cite{fn_real_part}. Thus, in the time domain we have
\begin{equation}
\widetilde{C}^{AA}_{ij}(t) = \int_{-\infty}^{\infty}
\widetilde{C}^{AA}_{ij}(\omega) \, cos(\omega t)\, d\omega \, ,
\end{equation}
which is even with respect to time inversion and can be Taylor
expanded in the following way
\begin{equation}
\widetilde{C}_{ij}^{AA}(t) = \sum_{n=0}^{\infty} \, (-1)^{n} \,
\frac{\mu_{ij}^{A,2n}}{(2n!)} \, t^{2n} \, ,
\label{eq:moment_expansion}
\end{equation}
with the moments given by
\begin{equation}
\mu_{ij}^{A,2n} = \int_{-\infty}^{\infty} \, \omega^{2n} \,
\widetilde{C}_{ij}^{AA}(\omega) \, d\omega \, .
\label{eq:moments}
\end{equation}
The next key step is to prove that these moments can be written as
certain correlation functions at $t = 0$, and can thus be computed as
regular thermal averages. By successive partial integration, one can
see that
\begin{equation}
\begin{split}
\widetilde{C}_{ij}^{AA}(\omega) & = \omega^{-2} \,
\widetilde{C}_{ij}^{A^{(1)}A^{(1)}}(\omega) = -\omega^{-2} \,
\widetilde{C}_{ij}^{A^{(2)}A}(\omega) \\ & =
\frac{(-1)^{m+n}}{\omega^{2m+2n+2p}} \,
\widetilde{C}^{A^{(2m+p)}A^{(2n+p)}}_{ij}(\omega) \, ,
\end{split}
\label{eq:C_derivative}
\end{equation}
where $A^{(n)}_{i}$ is the $n$-th time derivative of $A_{i}$. These
identities allow us to rewrite Eq.~(\ref{eq:moments}) in different
ways. From a computational viewpoint, it is convenient to use
\begin{equation}
\mu_{ij}^{A,2n} = \int_{-\infty}^{\infty}
\widetilde{C}^{A^{(n)}A^{(n)}}_{ij}(\omega) \, d\omega = \langle
A_{i}^{(n)}(0) A_{j}^{(n)}(0) \rangle \, ,
\label{eq:moment_static}
\end{equation}
which involves time derivatives of the lowest possible order. Finally,
the derivatives can be computed using Hamilton's equation of motion
\begin{equation}
\begin{split}
&\frac{d A}{d t} = \frac{\partial A}{\partial t} + \{ A, H \} \, ,\\
&\{ A, H\}  = \sum_{i} \left( \frac{\partial A}{\partial x_{i}}
\frac{\partial H}{\partial p_{i}} - \frac{\partial A}{\partial p_{i}}
\frac{\partial H}{\partial x_{i}} \right) \, .
\end{split}
\end{equation}
The Hamiltonian is
\begin{equation}
H= \sum_i \frac{ {p}_{i}^{2}}{ 2m_{i}}  + V(\{x_i \}) \, ,
\end{equation}
where $m_{i}$, $x_{i}$, and $p_{i}$ are mass, position and momentum,
respectively, and $V$ is a velocity-independent potential. It is
convenient to introduce $x_{i}'$ = $\sqrt{m_{i}}x_{i}$ to get rid of
the mass dependence in the kinetic energy. We will thus work with the
moments $\mu^{A',2n}_{ij}$ of the correlation function
$\widetilde{C}_{ij}^{A'A'}(t)$ =
$\sqrt{m_{i}m_{j}}\widetilde{C}_{ij}^{AA}(t)$, where $A_{i}'$ =
$\sqrt{m_{i}} A_{i}$.  The simplest object that gives information
about the vibrational spectrum is the position-position correlation
function. By taking $A_{i}'$ = $\bar{x}_{i}'$ = $x_{i}'-\langle x_{i}'
\rangle$, we obtain
\begin{equation}
\begin{split}
\mu^{\bar{x}',0}_{ij} & = \sqrt{m_{i}m_{j}} \, \left(\langle x_{i}
x_{j} \rangle - \langle x_{i} \rangle \langle x_{j} \rangle\right) \,
, \\
\mu^{\bar{x}',2}_{ij} & = \beta^{-1} \delta_{ij} \, , \\
\mu^{\bar{x}',4}_{ij} &= \frac{1}{\sqrt{m_{i}m_{j}}} \, \left \langle
\frac{\partial V}{\partial x_{i}} \frac {\partial V}{\partial x_{j}}
\right \rangle \, , \\
\mu^{\bar{x}',6}_{ij} &= \frac{ \beta^{-1}}{\sqrt{m_{i}m_{j}}} \sum_{k} \,
\frac{1}{m_{k}} \, \left \langle \frac{\partial^2
  V}{\partial x_{i} \partial x_{k} } \frac{\partial^2 V}{\partial
  x_{k} \partial x_{j} } \right \rangle \, ,
\end{split}
\label{eq:moments_x}
\end{equation}
for the four lowest moments of Eq.~(\ref{eq:moment_expansion}). Here
we reversed the mass scaling at the last step of the derivation, so as
to express the moments in terms of the regular atomic positions.  We
also used $\langle p_i' p_j '\rangle$ = $k_{\rm B}T\delta_{ij}$ =
$\beta^{-1} \delta_{ij}$ and $\langle
f(\{p_{i}'\})g(\{x_{i}'\})\rangle$ = $\langle f(\{p_{i}'\})\rangle
\langle g(\{x_{i}'\})\rangle$, where $f$ and $g$ are arbitrary
functions. Let us stress that this procedure renders
$\mu^{\bar{x}',2}_{ij}$ proportional to the identity matrix, a fact
that will be advantageous later. As an example of the freedom we have
in writing the moments, note that alternatively we can get
\begin{equation}
\mu_{ij}^{\bar{x}',4} = - \langle {x}_{i}'^{(3)} {x}_{j}'^{(1)}
\rangle = \frac{ \beta^{-1}}{\sqrt{m_{i}m_{j}}} \left\langle
\frac{\partial^2 V}{\partial x_{i} \partial x_{j} } \right\rangle \, .
\label{eq:dynamical-matrix}
\end{equation}
It is also interesting to note that other time-correlation functions
can be readily computed from Eq.~(\ref{eq:moments_x}). Indeed, it can
be seen from Eqs.~(\ref{eq:moments}) and (\ref{eq:C_derivative}) that
\begin{equation}
\mu_{ij}^{A^{(1)},2n} = \mu_{ij}^{A,2n+2} \, .
\end{equation}
Thus, for example, the lowest non-diagonal moment of the
velocity-velocity correlation function is $\mu_{ij}^{v',2} =
\mu_{ij}^{\bar{x}',4}$.

In order to gain physical insight, consider the case of a single
harmonic oscillator. There, we know the exact form of
$C^{\bar{x}'\bar{x}'}(\omega)$ = $\mu^{\bar{x}',0}[\delta(\omega -
  \omega_{0})+\delta(\omega + \omega_{0})]/2$, and from
Eqs.~(\ref{eq:moments}) and (\ref{eq:moments_x}) we obtain
$\omega_{0}^{2n} = \mu^{\bar{x}',2n} / \mu^{\bar{x}',0}$; then, for
$n=1$ we have $m \, \omega_{0}^{2}$ = $[\beta(\langle {x}^{2} \rangle
  - \langle x \rangle^{2})]^{-1}$. We thus find that the fluctuations
of the position give us the frequency of the (phonon) peak in the
$C^{\bar{x}'\bar{x}'}(\omega)$ spectrum. (This is essentially the
quasi-harmonic result used in
Refs.~\onlinecite{book-dove05,goodwin05}.) Similarly, for
$C^{v'v'}(\omega)$ we get $m\,\omega_{0}^{2}$ = $\langle
\partial^{2}V/\partial x^{2}\rangle$, i.e., the frequency is given by
the thermal-averaged dynamical matrix in this case. These intuitive
relations, which are exact in the harmonic limit and are generalized
below to the many-body case, are implicitly underlying the heuristic
methods to estimate phonon frequencies mentioned above
\cite{hellman11,brooks95,wheeler03}.

{\sl Practical scheme}.-- In general we will have many interacting
atoms and an anharmonic potential $V$. To simplify the problem, let us
make a unitary coordinate transformation $A_{a}' = \sum_i T_{ai}^{*}
A_{i}'$ that will be analogous to the usual change into a normal-mode
basis. We write the transformed correlation function and moments as
\begin{eqnarray}
 \widetilde{C}^{A'A'}_{ab}(t) & = & \sum_{ij} T^{*}_{ai} \widetilde
 C^{A'A'}_{ij}(t) T_{bj} \, , \\
\mu^{A',2n}_{ab} & = & \sum_{ij} T^{*}_{ai} \, \mu^{A',2n}_{ij} \, T_{bj} \, .
\label{eq:moment_transformation}
\end{eqnarray}
We choose $T_{ai}$ to diagonalize the lowest non-diagonal moment
$\check{\mu}_{ij}^{A'}$, so that
\begin{equation}
\sum_{ij} T^{*}_{ai} \, \check{\mu}^{A'}_{ij} \, T_{bj} =
\check{\mu}^{A'}_{a} \delta_{ab} \, .
\label{eq:transformation}
\end{equation}
Note that $T_{ai}$ gives the {\em polarization vectors} of our normal
modes. We then approximate \cite{fn_evenness}
\begin{equation}
\begin{split}
\widetilde{C}^{A'A'}_{ij}(t) & = \sum_{ab} T_{ia}^{*} \,
\widetilde{C}^{A'A'}_{ab}(t) \, T_{jb} \\ & \approx \sum_{a} T_{ia}^{*} \,
C^{A'A'}_{aa}(t) \, T_{ja}\, ,
\end{split}
\label{eq:decoupling}
\end{equation}
where the second equality is exact to low order in the moment
expansion. Hence, to investigate the spectrum given by
$\widetilde{C}_{ij}^{A'A'}(\omega)$, we will work with the collection of
anharmonic oscillators $C_{aa}^{A'A'}(\omega)$ \cite{fn_evenness}.

Now, we have $\check{\mu}^{\bar{x}'}_{ij}$ = $\mu^{\bar{x}',0}_{ij}$
and $\check{\mu}^{v'}_{ij}$ = $\mu^{v',2}_{ij}$. Interestingly, thanks
to the mass-scaling transformation, $\mu^{\bar{x}',2}_{ij}$ and
$\mu^{v',0}_{ij}$ are proportional to the identity matrix and will
remain diagonal in our normal-mode basis. Thus, since the two
lowest-order moments are diagonal for both
$\widetilde{C}_{ab}^{\bar{x}'\bar{x}'}$ and
$\widetilde{C}_{ab}^{v'v'}$, it seems reasonable to propose the
following {\em effective-harmonic approximation}
\begin{equation}
\begin{split}
C^{A'A'}_{aa}(t) & = \mu^{A',0}_{aa} \left( 1 - \frac{1}{2}
\frac{\mu^{A',2}_{aa}}{\mu^{A',0}_{aa}} t^{2}+ \ldots \right) \\ &
\approx \mu^{A',0}_{aa} \cos(\omega_{a} t) \, ,
\label{eq:effective_harmonic}
\end{split}
\end{equation}
where $\omega_a = \sqrt{ \mu^{A', 2}_{aa}/ \mu^{A', 0}_{aa}}$.

For $A_{i}'$~=~$\bar{x}_{i}'$, Eq.~(\ref{eq:transformation}) involves
the diagonalization of $\mu_{ij}^{\bar{x}',0}$ = $ \sqrt{m_{i}m_{j}}
(\langle x_{i}x_{j}\rangle -\langle x_{i} \rangle \langle x_{j}
\rangle)$; for $A_{i}'$~=~$v_{i}'$ we diagonalize $\mu_{ij}^{v',2}$ =
$\mu_{ij}^{\bar{x}',4}$, which can be expressed as the
thermal-averaged dynamical matrix of
Eq.~(\ref{eq:dynamical-matrix}). Hence, the eigenvalues of these
matrices provide us with a rigorously justified approximation to the
position of the (phonon) peaks in the $C_{aa}^{A'A'}(\omega)$ spectra;
as described above, they render the exact phonon frequencies in the
harmonic limit.

In order to capture more complex line shapes, we need a strategy to
treat the full $C_{aa}^{A'A'}(\omega)$. One possibility is to assume an
analytic form for it, with free parameters that can typically be
written as a function of the low-order moments. We worked with some
physically motivated choices, i.e., Gaussians, Lorentzians, and
combinations of delta functions. Such an approach allowed us to
compute the peak frequencies for a variety of model systems, even in
the presence of significant non-linear effects (i.e.,
overtones). However, the scheme failed to provide a quantitative
estimate of the peak widths; further, for models with very strongly
$T$-dependent frequencies, we sometimes obtained non-physical
solutions for the parameters of the trial spectral functions.

Fortunately, we found it possible to treat $C_{aa}^{A'A'}(\omega)$ in a
robust and accurate way by resorting to Mori's continued-fraction
representation \cite{mori65}. Using the notation of
Ref.~\onlinecite{cuccoli92}, we write for a generic, real
autocorrelation function $C(\omega)$
\begin{equation}
\begin{split}
& C(\omega) = \frac{\mu^0}{\pi} \Re[\psi_0(i \omega)]\, , \\
& \psi_n(z) = \frac{1}{ z+ \delta_{n+1}\psi_{n+1}( z) } \, ,
\end{split}
\end{equation}	
where the $\delta_{n}$ parameters are explicit functions of the
moments. The first three are given by \cite{cuccoli92}
\begin{equation}
\delta_1 = \frac{\mu^2 }{ \mu^0},\; \delta_2 = \frac{\mu^4}{\mu^2}
-\frac{\mu^2 }{ \mu^0},\; \delta_3 = \frac{1}{\delta_2} \left [
  \frac{\mu^6}{\mu^2} - \left [ \frac{\mu^4}{\mu^2} \right]^2 \right]
\, .
\end{equation}
A continued fraction is usually terminated by assuming the last
$\psi_{n}(z)$ term to be the Laplace transform of some model
function. Many terminations have been used in the literature
\cite{cuccoli92,cuccoli93,cowley94,macchi96,balucani03}; yet, we found
that, for the model systems we investigated, the line shape is quite
insensitive to the termination scheme when including up to sixth
moments; thus, we simply used for $\psi_{2}(z)$ the Gaussian
termination described in Ref.~\onlinecite{cuccoli92}. Obviously, in
this scheme a separate continued-fraction expansion must be calculated
for each $C^{A'A'}_{aa}(\omega)$. In practice, we first compute the {\em
  atomic} moments $\mu_{ij}^{A',2n}$, and then obtain those
corresponding to our normal modes by using
Eq.~(\ref{eq:moment_transformation}).

{\sl Example of application}.-- To test our approach, we used it to
compute vibrational spectra -- with moments calculated from MC
simulations -- and compared the results with the exact ones obtained
from MD \cite{fn_methods,allen89}. We worked with model systems, which
gave us full control of the potential-energy surface and allowed us to
try the method in very diverse and challenging situations. Overall we
found that our approach renders excellent results for the main
features of the vibrational spectrum. To illustrate the method, here
we describe a particularly demanding case, namely, a system undergoing
a structural phase transition driven by a soft phonon mode.

Let us consider a cubic crystal with three degrees of freedom
$x_{l\alpha}$ {\sl per} cell $l$, where $\alpha$ = $x$, $y$, and
$z$. For simplicity we take $m = 1$ and write the Hamiltonian as
\begin{equation}
\begin{split}
H = & \;  \frac{1}{2} \sum_{l\alpha} p_{l\alpha}^2 + c_{1} \sum_{l}
(|\vec{x}_{l}|^{2}-1)^{2} \; + \\
& c_{2} \sum_{l} (x_{lx}^{2}x_{ly}^{2} + x_{ly}^{2}x_{lz}^{2} +
x_{lz}^{2}x_{lx}^{2}) \; + \\
& \frac{1}{2}\sum_{ll^{'}}{}^{'} \left[ c_{3}
  |\vec{x}_{l}-\vec{x}_{l^{'}}|^{2} + c_{4} 
  |x_{l\alpha(ll')}-x_{l'\alpha(ll')}|^{2} \right] \, ,
\end{split}
\end{equation}
where the primed sum is restricted to nearest-neighboring cells. In
essence this is the well-known discrete $\phi^{4}$
model~\cite{rubtsov00,fnmeanfield,bruce80}, extended to include (1) an
on-site anisotropic term ($c_{2}>0$) chosen so that the ground state
has a tetragonal symmetry and (2) a coupling between nearest-neighbors
($c_{4}$) that breaks the symmetry between longitudinal and
transversal phonon branches. Here we show representative results
obtained for a choice of parameters ($c_{1}$~=~$0.25$,
$c_{2}$~=~$0.50$, $c_{3}$~=~$1.00$, $c_{4}$~=~$0.50$) that leads to a
second-order displacive \cite{fnorderdisorder} phase transition. While
the energy units are arbitrary, this model renders a realistic
representation of a phase transition at room temperature.

We simulated the model in a periodically-repeated
20$\times$20$\times$20 supercell. We carefully checked the convergence
of the MC and MD simulations. For example, the MD results shown here
were obtained by Fourier transforming time-correlation functions
computed from constant-energy trajectories whose starting points were
snapshots taken from a constant-$T$ Langevin simulation. For each $T$
investigated, ten different starting points were considered, the final
spectral function being an average.

In periodic systems the moments, as well as the time-correlation and
spectral functions, become block-diagonal in the Bloch
representation. Hence, instead of the $\widetilde{C}_{ij}^{A'A'}$ and
$C_{aa}^{A'A'}$ functions in the formulas above, in the following we
will use the Fourier-transformed
$\widetilde{C}_{\mathbf{q},\alpha\beta}^{A'A'}$ and
$C_{\mathbf{q},aa}^{A'A'}$, where $\mathbf{q}$ is a vector in the
first Brillouin zone of our model crystal; the corresponding moments
are $\mu_{\mathbf{q},\alpha\beta}^{A',2n}$, etc.

\begin{figure}[t!]
 \includegraphics[width=0.9\columnwidth]{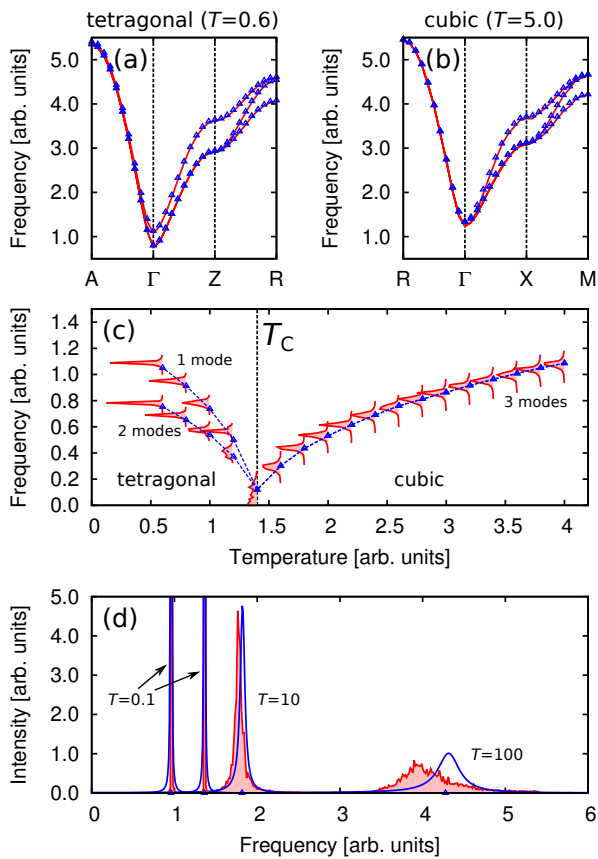}
 \caption{(Color online.) Dynamical properties of our model system. We
   show the results obtained from our approach based on MC simulations
   (blue) and the {\em exact} MD results (red). Panels~(a) and (b):
   Phonon bands at two different $T$'s. Panel~(c): Spectrum at the
   $\Gamma$ point, for $T$'s around the critical temperature $T_{\rm
     C}$. Panel~(d): $\Gamma$ spectrum at selected $T$'s. All spectra
   are normalized to unity. See text for more details.}
\label{fig:results}
\end{figure}

Figures~\ref{fig:results}(a) and \ref{fig:results}(b) show the phonon
bands at low and high $T$'s, respectively. The MD results were
obtained from the first moment of the peaks in the spectrum given by
the trace of
$\widetilde{C}_{\mathbf{q},\alpha\beta}^{v'v'}(\omega)$. The MC
results were obtained by diagonalizing
$\mu_{\mathbf{q},\alpha\beta}^{v',2}$ within our effective-harmonic
approximation [Eq.~(\ref{eq:effective_harmonic})].  The agreement is
excellent.

Figure~\ref{fig:results}(c) shows the $T$-dependent frequencies of the
phonons at the $\Gamma$ point ($\mathbf{q} = 0$). In the cubic phase,
the frequencies decrease as $T$ is reduced, and essentially vanish at
the critical temperature $T_{\rm C}$. Then, below $T_{\rm C}$ the
frequencies increase as $T$ decreases. We have three $\Gamma$ phonons
in both phases: they are three-fold degenerate in the cubic structure,
but split in two groups when the symmetry is lowered to tetragonal. In
the figure we show the trace of the
$\widetilde{C}_{\Gamma,\alpha\beta}^{\bar{x}'\bar{x}'}(\omega)$
functions resulting from MD simulations, as well as the frequencies
computed at the effective-harmonic level by diagonalizing
$\mu_{\Gamma,\alpha\beta}^{\bar{x}',0}$ obtained from MC. This
approximation gives excellent results, even in the immediate vicinity
of $T_{\rm C}$ where the system is strongly anharmonic.

Figure~\ref{fig:results}(d) shows the line shapes from the
continued-fraction representation of $C_{\Gamma,aa}^{\bar{x}'\bar{x}'}
(\omega)$, using up to the sixth-order moments, together with the MD
results. We can appreciate that the widths of the peaks obtained with
our method are semi-quantiatively correct. We include a result at an
unrealistically high $T = 100$, where the peak broadening is very
significant. Even in such extreme conditions, our approximate spectral
function provides a fair representation of the exact one.

{\sl Final remarks}.-- We have shown that the main features of
classical vibrational spectra can be accurately computed from
knowledge of the low-order moments of the appropriate time-correlation
functions. More precisely, we have presented a way to compute a
normal-mode-like basis that renders the low-order moments diagonal,
which allows us to approximate the full spectrium by a collection of
anharmonic oscillators [Eq.~(\ref{eq:decoupling})]. Further, we have
introduced an {\em effective-harmonic approximation}
[Eq.~(\ref{eq:effective_harmonic})] that makes it possible to obtain
very accurate results for the vibrational frequencies from simple
statistical averages of atomic positions or forces. We have also shown
that it is possible to reproduce the line shape of the spectral
functions in a semi-quantitative way, provided higher moments are
available.

The moments can be obtained as thermal averages from MC
simulations. Alternatively, one may obtain them from MD simulations,
without the need to explicitly compute the time-correlation functions;
this should allow for shorter MD runs (only as long as needed to
compute accurate thermal averages) and simplify the use of thermostats
(as their interfering with the dynamics would be unimportant).

Our effective-harmonic treatment provides a rigorous justification for
some of the assumptions underlying previous schemes in the literature
\cite{hellman11,brooks95,wheeler03}. Further, first-principles methods
like the one proposed in Ref.~\onlinecite{hellman11} could greatly
benefit from results such as the identity $\langle \partial^{2}V
/\partial x_{i}\partial x_{j} \rangle$ = $\beta \langle (\partial V /
\partial x_{i}) (\partial V / \partial x_{j})\rangle$, which we have
proven here and constitutes a conveneint way to obtain the
thermal-averaged force-constant matrix (computationally very costly)
from appropriate products of forces (readily available). Finally, our
effective-harmonic approximation can be connected with quasi-harmonic
methods that have been applied in a variety of
contexts~\cite{book-dove05,goodwin04,goodwin05}; our results support
the applicability of such schemes even in cases with significant
anharmonicity.

We hope the methods here discussed will become standard tools in
classical simulations, where they can be used to a great advantage.

Work supported by the EC-FP7 project OxIDes (Grant No. CP-FP 228989-2)
and MINECO-Spain (Grants No. MAT2010-18113, No. MAT2010-10093-E, and
No. CSD2007-00041). Discussions with J.C.~Wojde\l\ are gratefully
acknowledged.


\begin{thebibliography}{10}
\bibitem{fn_evenness}%
  \BibitemOpen
  \bibinfo {note} {We can write the autocorrelation functions $C_{aa}^{AA}(t)$
  and $C_{ii}^{AA}(t)$ without a tilde because they are even, as the
  corresponding spectral functions $C_{aa}^{AA}(\omega)$ and
  $C_{ii}^{AA}(\omega)$ are real.}%
  \bibAnnoteFile{Stop}{fn_evenness}%
\bibitem{fn_methods}%
  \BibitemOpen
  \bibinfo {note} {The statistical methods we used are standard. A description
  can be found, for example, in Ref.~\protect\onlinecite{allen89}.}%
  \bibAnnoteFile{Stop}{fn_methods}%
\bibitem{fnmeanfield}%
  \BibitemOpen
  \bibinfo {note} {This kind of models have been studied extensively (see
  Ref.~\protect\onlinecite{bruce80} for a review) and solved in approximate
  ways. Our method might seem related with some of the mean-field schemes
  discussed in the literature, such as the so-called independent-mode
  approximation. However, that resemblance is misleading, as we do simulate the
  true equilibrium state of the material, and introduce approximations only to
  extract the dynamical information.}%
  \bibAnnoteFile{Stop}{fnmeanfield}%
\end{thebibliography}
\end{document}